\documentclass[12pt,preprint]{aastex}
\pdfoutput=1
\newcommand\gsim{\,\lower3pt\hbox{$\sim$}\llap{\raise2pt\hbox{$>$}}\,}
\newcommand\lsim{\,\lower3pt\hbox{$\sim$}\llap{\raise2pt\hbox{$<$}}\,}

\begin{document}

\title{The Emergence of a Twisted Flux Tube into the Solar Atmosphere:
Sunspot Rotations and the Formation of a Coronal Flux Rope}

\author{Y.~Fan}
\affil{HAO, ESSL, National Center for Atmospheric
Research\altaffilmark{1}, P.O. Box 3000, Boulder, CO 80307}

\altaffiltext{1} {The National Center for Atmospheric Research
is sponsored by the National Science Foundation}

\begin{abstract}
We present a 3D simulation of the dynamic emergence of a twisted magnetic
flux tube from the top layer of the solar convection zone into the solar
atmosphere and corona. It is found that after a brief initial stage of flux 
emergence during which the two polarities of the bipolar region become
separated and the tubes intersecting the photosphere become
vertical, significant rotational motion sets in
within each polarity.  The rotational motions of the two polarities are
found to twist up the inner field lines of the emerged fields such that
they change their orientation into an inverse configuration (i.e. pointing
from the negative polarity to the positive polarity over the neutral line).
As a result, a flux rope with sigmoid-shaped, dipped core fields form in
the corona, and the center of the flux rope rises in the corona
with increasing velocity as the twisting of the flux rope footpoints continues.
The rotational motion in the two polarities is
a result of propagation of non-linear torsional Alfv\'en waves along the flux tube,
which transports significant twist from the tube's interior portion towards
its expanded coronal portion.  This is a basic process whereby twisted
flux ropes are developed in the corona with increasing twist and magnetic
energy, leading up to solar eruptions.
\end{abstract}


\section{Introduction}

How twisted magnetic fields emerge from the dense, convectively
unstable solar convection zone into the stably stratified, rarefied
solar atmosphere and corona is a fundamentally important question for
understanding the formation of solar active regions and the development
of precursor structures for solar eruptions such as flares and CMEs.
Pioneered by the earlier work of Shibata and collaborators
\citep{shibataetal1989}, a large body of 3D MHD simulations of the emergence
of twisted magnetic flux tubes into the solar atmosphere
\citep[e.g.][]{fan2001, magara_longcope2001,
magara_longcope2003, magara2004, manchestetal2004, archontisetal2004,
murrayetal2006, magara2006, archontis_toeroek2008} have been carried out
in recent years, which have provided important insight into the above
dynamic process.
It has been shown that magnetic flux reaching the photosphere can
undergo a dynamic expansion into the atmosphere as a result of the
non-linear growth of the magnetic buoyancy instability.
The conditions required of the emergent tubes for the onset
of such dynamic expansion into the atmosphere has been examined by 
\citet{archontisetal2004} and \citet{murrayetal2006}.

These simulations have also shown that it is difficult for a twisted flux
tube to rise bodily into the corona as a whole due to the heavy plasma
that is trapped at the bottom concave (or U-shaped) portions of the
winding field lines \cite[e.g.][]{fan2001, manchestetal2004,
archontisetal2004, magara2006}.  While the upper parts of the helical
field lines of the twisted flux tube expand into the atmosphere and the
corona, the U-shaped parts of the winding field lines remain largely
trapped at and below the photosphere layer, and the center of
the original tube axis ceases to rise a couple of pressure scale heights
above the photosphere.  Despite this, a twisted coronal flux rope with
sigmoid-shaped concave upturning field lines threading under a central
axis that rises into coronal heights is found to develop in the
end in many simulations
\citep[e.g.][]{magara_longcope2001, magara_longcope2003, magara2004,
manchestetal2004, magara2006, archontis_toeroek2008}.
\citet{manchestetal2004} found that shear flows on the photosphere
driven by the magnetic tension force continually transport axial flux
and magnetic energy to the upper expanded portion of the emerged
field, causing it to eventually ``pinch off'' from the lower mass-laden
part via magnetic reconnection in a current sheet forming in the
lower atmosphere. This reconnection produces a coronal flux rope with a
new central axial field line that rises into the corona with increasing
speed.  Such development of a coronal flux rope is found to take
place more readily with a more sharply arched subsurface tube for which the
emerging segment breaking through the photosphere contains fewer turns
of field line twist.
\citet{magara_longcope2003} shows that during the emergence of a twisted
flux tube, both the vertical emergence of magnetic flux and the horizontal
motions at the footpoints of the emerged field lines inject magnetic energy
and helicity (or twist) into the atmosphere. The contribution by the direct
emergence dominates at the early phase of flux emergence, while the
horizontal flows become a dominant contributor of helicity injection later.
\citet{magara2006} studied in more detail the horizontal flows on
the photosphere during flux emergence and found that as soon as the flux tubes
intersecting the photosphere become vertical and the two polarity flux
concentrations on the photosphere become separated, a rotational flow appears
in each of the polarity flux concentrations.

In the present paper, we study further the role of the shear flows and 
the rotational flows in driving the formation of a coronal flux rope
during flux emergence.  We present a simulation of the emergence of
a twisted $\Omega$-shaped subsurface flux tube whose emerging
segment initially breaking through the photosphere
contains less than one full wind of field-line twist.
As in the simulation of \citet{magara2006}, we find a brief initial phase
of flux emergence during which the two polarity flux concentrations on
the photosphere undergo a shearing motion along the neutral line and become
separated. Then a prominent rotational/vortical motion develops within
each polarity, reminiscent of sunspot rotations that have been observed
occasionally \citep[e.g.][]{brownetal2003}.
The rotational motions of the two polarities are due to torsional
Alfv\'en waves propagating along the twisted flux tube, transporting
magnetic twist from the tube's interior portion, where the rate of
twist is high, to its greatly stretched coronal portion, where the
rate of twist is low, a result that was predicted by an earlier idealized
analytical model of \citet{longcope_welsch2000}.
The rotational motions of the two polarities are found to persist
throughout the later phase of the evolution and steadily
inject magnetic helicity into the atmosphere.
The combined effect of the initial shear flow and the subsequent
rotational motions of the two polarities is to twist up the
inner field lines of the emerged field,
causing them to rotate from their initial normal configuration (i.e. arching
over the polarity inversion line from the positive to the negative
polarity), into an inverse configuration (directed from the negative
polarity to the positive polarity over the neutral line).
In this manner, a flux rope with sigmoid-shaped dipped core fields forms
in the atmosphere, with the center of the flux rope, as represented by the
O-point of the transverse field in the central cross-section, rising into
coronal heights.
This field reconfiguration is a major new finding of our study, namely
that the horizontal rotation of the field direction of the lower
emerged flux into an inverse configuration causes an upward propagation
of the O-point, and initiates the development of the coronal flux rope.
Similar to many previous simulations \citep[e.g.][]{magara_longcope2003,
magara2004, manchestetal2004, magara2006}, a current sheet
forms in the lower atmosphere, and the field lines whose lower parts going
through the current sheet all display a sigmoid morphology, which may
correspond to the observed X-ray sigmoid loops in active regions.

The remainder of the paper is structured as follows. In \S 2 we describe
the numerical MHD model and the detailed setup of the simulation.
In \S 3, we describe the simulation results, with focuses on examining the
nature of the rotational motion of the two polarities on the photosphere and
how it drives the formation of a flux rope in the corona.
Finally in \S 4 we summarize the conclusions and discuss the implications
of our simulation results.

\section{The numerical model}
In this study we solve numerically the following ideal magnetohydrodynamic
(MHD) equations in a Cartesian domain:
\begin{equation}
{\partial \rho \over \partial t}
+ \nabla \cdot ( \rho {\bf v}) = 0 ,
\label{eqcont}
\end{equation}
\begin{equation}
{\partial ( \rho {\bf v} ) \over \partial t}
+ \nabla \cdot ( \rho {\bf v} {\bf v} )
= - \nabla p - \rho {\bf g} + {1 \over 4 \pi}
( \nabla \times {\bf B} ) \times {\bf B}
\label{eqmom}
\end{equation}
\begin{equation}
{\partial {\bf B} \over \partial t}
= \nabla \times ({\bf v} \times {\bf B}),
\label{eqinduc}
\end{equation}
\begin{equation}
\nabla \cdot {\bf B} = 0 ,
\label{eqdivb}
\end{equation}
\begin{equation}
{\partial e \over \partial t} = - \nabla \cdot
\left [ \left ( \varepsilon + {\rho v^2 \over 2} + p \right ) {\bf v}
- {1 \over 4 \pi} ({\bf v} \times {\bf B} ) \times {\bf B} \right ]
+ \rho {\bf v} \cdot {\bf g},
\label{eqetot}
\end{equation}
\begin{equation}
p = {\rho R T \over \mu},
\label{eqtherm}
\end{equation}
where
\begin{equation}
e=\varepsilon+ {\rho v^2 \over 2} + {B^2 \over 8 \pi} ,
\end{equation}
\begin{equation}
\varepsilon = {p \over {\gamma - 1} } .
\end{equation}
In the above ${\bf v}$, ${\bf B}$, $\rho$, $p$, $T$, $R$, $\mu$,
$\varepsilon$, $e$ denote
respectively the velocity field, the magnetic field, density,
pressure, temperature, the gas constant, the mean molecular
weight, the internal energy per unit volume, and the total
energy (internal+kinetic+magnetic) per unit
volume. The gravitational acceleration
${\bf g} = - g {\hat {\bf z}}$ is a constant and is directed
in the $-z$ direction.
The above ideal MHD equations are solved numerically without
any {\it explicit} viscosity, magnetic diffusion, and non-adiabatic
effects, although numerical dissipations are present.
Thus we do not physically treat the radiative energy transfer,
thermal conduction, coronal heating, and the effects of partial
ionization, which all play important roles in the thermal energy
evolution of the plasma in the solar atmosphere.
The thermal properties of the plasma in our simulations
are therefore unrealistic.
Our study here focuses on the dynamic and topological
properties of the 3D emerging magnetic field, given the
above simplifications.

The basic numerical schemes we use to solve the above MHD
equations are as follows.  The equations are discretized in
space using a staggered finite-difference scheme
\citep{stone_norman1992a}, and advanced in time with an
explicit, second order accurate, two-step predictor-corrector
time stepping.
A modified, second order accurate Lax-Friedrichs scheme similar
to that described in \citet[][see eq. (A3) in that paper]
{rempeletal2008} is applied for evaluating
the fluxes in the continuity, momentum, and energy equations. 
Compared to the standard second order Lax-Friedrichs scheme,
this scheme significantly reduces numerical diffusivity for
regions of smooth variation, while retaining the same
robustness in regions of shocks.
A method of characteristics that is upwind in the Alfv\'en
waves \citep{stone_norman1992b} is used for evaluating the
${\bf v} \times {\bf B}$ term in the induction equation, and
the constrained transport scheme is used to ensure $\nabla 
\cdot {\bf B} = 0$ to machine round-off errors.

The equations are non-dimensionalized
using the following normalizing constants:
$\rho_{\rm ph} \approx 2.7 \times 10^{-7} $ g ${\rm cm}^{-3}$
(photospheric density),
$H_{\rm ph} \approx 150 $ km (pressure scale height at the photosphere),
$T_{\rm ph} \approx 5100$ K (photospheric temperature),
$v_{\rm th} = ({\cal R} T_{\rm ph} / \mu)^{1/2} \approx 6.5 $ km~${\rm s}^{-1}$
(isothermal sound speed at the photosphere), $t_{\rm ph} = H_{\rm ph}
/ v_{\rm th} \approx 23$ s,
$p_{\rm ph} = \rho_{\rm ph} v_{\rm th}^2 $ (photospheric pressure)
and $B_{\rm ph} = (4 \pi \rho_{\rm ph} v_{\rm th}^2)^{1/2} \approx 1200$ G,
as the units for density, length, temperature, velocity, time, pressure,
and magnetic field respectively.
For the remainder of the paper, quantities and equations are expressed in
the above units unless otherwise specified.

The setup of the numerical simulation is illustrated in
Figure 1. The simulation domain (top panel of Figure 1) is
$(-69.3, -63.6, -20) \leq (x, y, z) \leq (69.3, 63.6, 115)$,
resolved by a $320 \times 300 \times 392$ grid.
In the $x$-direction ($y$-direction), the mesh size is
$\Delta x = 0.2$ ($\Delta y = 0.2$) within
$-20 \leq x \leq 20$ ($-20 \leq y \leq 20$), and it increases
gradually for $|x| > 20$ ($|y| > 20$), reaching about 2 at the boundaries.
In the $z$-direction, the mesh size is $\Delta z = 0.2$ within the
range $-20 \leq z \leq 32$, and for $z > 32$ it increases gradually
to 1.39 at the top boundary.
The horizontal boundaries in the $x$-direction (the direction of
the initial flux tube) are assumed to be periodic.
The bottom boundary is a perfectly conducting wall and is non-penetrating.
An outflow boundary condition \citep[see][]{stone_norman1992a}
is applied for the top boundary and the two horizontal boundaries in
the $y$-direction.
For the initial state, we consider a plane-parallel atmosphere in
hydrostatic equilibrium, composed of an adiabatically
stratified polytropic layer representing the top layer of the convection
zone, an isothermal layer representing the photosphere and the
chromosphere, connecting to another isothermal layer of 1 MK representing
the corona.
The vertical profiles of the pressure $p_0 (z)$, density $\rho_0 (z)$,
and temperature $T_0 (z)$ of the hydrostatic field-free atmosphere
are shown in the bottom panel of Figure 1.
A horizontal, twisted magnetic flux tube oriented along the $x$-direction
is embedded in the polytropic layer (convection zone) with its axis located
at $(y,z) = (0, -14)$. The magnetic field of the flux tube is given by
\begin{equation}
{\bf B} = B_x (r) \, \hat{\bf x} + B_{\theta} (r) \, \hat {\bf \theta},
\end{equation}
where
\begin{equation}
B_x (r) = B_{0} \exp (- r^2 /a^2 ) ,
\end{equation}
\begin{equation}
B_{\theta} (r) = q r B_x (r) .
\end{equation}
In the above $\hat{\bf x}$ is the tube axial direction,
$\hat{\bf {\theta}}$ is the azimuthal direction in the tube cross-section,
and $r$ denotes the radial distance to the tube axis.
The flux tube is uniformly twisted, with the
constant $q$ denoting the angle of field line rotation about the
axis over a unit length of the tube.
We set $B_0 = 5$, the radius $a=2$, and the rate of twist
$q= - a^{-1}$ (left-handed), which makes the tube marginally stable
for the kink instability \citep{lintonetal1996}.
The profile of the magnetic pressure $p_m$ along the vertical
line at $(x,y) = (0,0)$ through the center of the embedded tube is also
shown in Figure 1.
Inside the flux tube, the plasma pressure differs from that of the
hydrostatic field-free atmosphere $p_0 (z)$ by:
\begin{equation}
p_1 (r) = - {B_0^2 \over 2} \exp (- 2 r^2 / a^2 )
\left [ 1 - {1 \over 2} q^2 a^2 \left ( 1 - {2 r^2 \over a^2}
\right ) \right ],
\label{p1}
\end{equation}
satisfying the equation:
\begin{equation}
{dp_1 \over dr} = -{d \over dr} \left ( {B_x^2+B_{\theta}^2 \over 2 } \right )
-{ B_{\theta}^2 \over r}.
\label{tubeeq}
\end{equation}
which ensures that the pressure gradient in the flux tube balances
the Lorentz force. To make the flux tube buoyant,
we further impose a density change $\rho_1 $ within the
flux tube relative to $\rho_0(z)$ of the field-free hydrostatic atmosphere:
\begin{equation}
\rho_1 = - \rho_0(z) {B_x^2 (r) \over 2 p_0(z)}
\left [ (1+\epsilon) \exp ( - x^2 / \lambda^2 ) - \epsilon \right ]
\end{equation}
where $\lambda = 8$ and $\epsilon = 0.2$. This makes the middle portion
of the flux tube buoyant, with $\rho_1 / \rho_0 = -0.11$ at the middle of
the tube axis. The buoyancy declines with $|x|$ from $x=0$
following a Gaussian profile with an e-folding length of $\lambda$,
and the two ends are slightly anti-buoyant with $\rho_1 / \rho_0 = 0.022$
at the ends of the tube axis.
The plasma $\beta$ (defined as the ratio of the plasma pressure over the
magnetic pressure) is about $8.3$ at the axis of the tube.

\section{Results}
\subsection{Overview of the evolution}
The early evolution of the buoyant flux tube is qualitatively similar to
those reported in many previous investigations \citep[e.g][]
{magara_longcope2001, fan2001, magara2004, manchestetal2004,archontisetal2004,
murrayetal2006,magara2006}.  The central portion of the initial flux tube
rises buoyantly while the two ends of the flux tube
sink down towards the bottom owing to the slight negative
buoyancy there.  The flux tube
develops into an $\Omega$-shape, whose central apex portion rises towards
the photosphere.  When the flux tube enters the stably stratified
photosphere, magnetic pressure builds up at the photosphere layer and 
the magnetic buoyancy instability develops. This causes a rapid expansion
of the front of the flux tube into the solar atmosphere and initiates the
flux emergence.

Figures 2 and 3 show respectively two perspective views of
the 3D magnetic field as delineated by a set of representative field lines
at 4 time instances during the flux emergence into the solar atmosphere.
We see that at $t=65$, the apex of the flux tube has broken through
the photosphere and the front of the emerging flux tube begins to
expand rapidly into the atmosphere and corona. With time, a twisted flux
rope structure with field lines winding about a central axis (as represented
by the black line) is formed in the atmosphere (see the $t=100$ and $t=120$
panels).  By time $t=120$, the central axis of the flux rope structure has
reached coronal heights, and sigmoid-shaped field lines (see the orange and
red field lines), some of which containing dips,
thread under the central axis with an
inverse configuration (i.e. pointing from negative to positive polarity region
over the neutral line).  Note that field lines in different images are not the
same field lines carrying the same plasma.  They are simply field lines selectively
drawn in each instance to reflect the topological structure of the 3D field.
%%The black axis in each panel is chosen to be the field line which, at that
%%instance, is parallel to the $x$-axis (i.e. with $B_y = 0$) at its intersection
%%with the central line of $x=0$ and $y=0$.
For example, the black axis in each panel corresponds to a {\it different}
field line in each of the time instances,
an important point to be addressed later in \S 3.2.

Figures 4(a)(b) show the evolution of the height and rise velocity for 3
different points within the central vertical plane of $x=0$: (1) the front
of the emerging flux tube, (2) the Lagrangian position of
the original tube axis, and (3) the O-point of the transverse magnetic field
in the central vertical plane at $x=0$, which is the point where $B_y = 0$
on the central vertical line of $x=0$ and $y=0$.
Figure 5 shows an example snapshot of the central vertical plane of $x=0$ at
time $t=108$, where the arrows show the direction of the transverse
magnetic field in the plane, overlaid with a color image showing
density on a log scale. Note, only the part above the photosphere is shown,
and the arrow lengths are all the same, normalized by the transverse field
strength, for the points where the transverse field strengths are non-zero.
The O-point of the transverse magnetic field is marked by the pink `*' point,
and it has reached the height of the base of the corona at $t=108$.
It should be emphasized that the O-point position is based on a geometric
definition which depends on the variation of $B_y$ with $z$ along the
central vertical line of $x=0$ and $y=0$. Thus it does not track a
Lagrangian plasma element, and its height variation with time
{\it does not in general reflect the true vertical plasma motion} but is
strongly influenced by the change with time of the horizontal magnetic
field orientations at points along the central vertical line.
For these reasons, $dz/dt$ of $z(t)$ for the O-point position
shown in Figure 4(a), does not in general agree with $v_z$,
the vertical velocity of the plasma instantaneously residing at the
O-point shown in Figure 4(b) (dashed-line).
We also note that the black field line shown in each of the panels in
Figure 2 and 3 is the field line that is drawn from the O-point
at that instance, and it corresponds to a different field line in each time.

From the position and the rise velocity of the front of the flux tube as shown
in Figure 4, one can see that the onset of the magnetic buoyancy instability
and flux emergence into the atmosphere takes place at about $t=60$, at
which time there is an exponential increase of the vertical speed of 
the front.  \citet{archontisetal2004} and \citet{murrayetal2006} have, based
on the original work of \citet{newcomb1961, thomas_nye1975} and \citet{acheson1979},
given the following critical condition for the onset of magnetic buoyancy instability
and flux emergence at the photospheric layer:
\begin{equation}
-H_p {\partial \over \partial z} \left ( \log B \right ) >
- {\gamma \over 2} \beta \delta + {k_\parallel}^2 \left (
1 + {{k_\perp}^2 \over {k_z}^2} \right ),
\end{equation}
where $z$ is height, $H_p$ is the local pressure scale height at
the photosphere, $B$ is the magnetic field strength, $\gamma$ is
the ratio of the specific heats, $\beta$ is the ratio of the plasma
pressure over the magnetic pressure, $\delta \equiv
\nabla - \nabla_{ad}$ is the superadiabaticity of the atmosphere, which
is $-0.4$ for the isothermal stratification of the photosphere, and
$k_{\parallel}$, $k_{\perp}$ and $k_z$ are the three components of the
local perturbation wave vector (normalized by $1/H_p$), with
$k_{\parallel}$ and $k_{\perp}$ being the two horizontal components
parallel and perpendicular to the local magnetic field direction, and
$k_z$ being the $z$ component.  We find that at $t=60$
the build-up of magnetic pressure at the photosphere
layer has just exceeded the gas pressure, forming a layer of low plasma $\beta$
(see top panel of Figure 6), where the strong stabilizing term $- (\gamma / 2)
\beta \delta$ is reduced to become smaller than $ -H_p (\partial / \partial z)
(\log B)$ (see bottom panel of Figure 6).  Using the horizontal widths $l_x$ and $l_y$
of the emerging magnetic region at $z=2$ to estimate
$k_{\parallel} \sim 2 \pi / ly$ and $k_{\perp} \sim 2 \pi / l_x$, and $k_z \sim
2 \pi /$ tube radius, we find that the above critical condition is met at $z \approx 2$.

As the front of the emerging flux tube begins to expand into the atmosphere at about
$t=60$, the rise speed of the original tube axis (solid curve in Figure 4(b)) and the
rise speed measured at the location of the O-point (dashed curve in Figure 4(b)) both
begin to decline, and their positions remain below $z=5$ during the period
from $t=65$ to $t=86$ (see Figure 4(a)).  
However, soon after $t=86$ there is an upward jump of the O-point position,
and then its height continues to increase sharply. With the O-point
moving into coronal heights, a flux rope with dipped, sigmoid-shaped core field
(see the $t=120$ panels in Figures 2 and 3) develops in the corona.

\subsection{Vortical motion of the two polarities and flux rope formation}

To understand the apparent rise of the O-point into coronal heights and the formation
of the coronal flux rope, we have tracked the evolution of a set of field lines which
intersect the center vertical line of the domain (at $x=0$ and $y=0$).  The way we do this
is by tracking the motion of a few Lagrangian points situated along the center vertical line
of symmetry at $x=0$ and $y=0$, and trace out field lines through these points at
successive time instances from $t=65$ to $t=120$.
Figures 7, 8, and 9 show 3D evolution of these tracked field lines as viewed from 3 perspectives.
The black field line here corresponds to the original tube axis, and all the other field
lines have their mid points (at $x=0$, $y=0$) above the mid point of the black
field line (reddish field lines have mid points above bluish field lines).
Note that the apex of the red field line does not correspond
to the emerging front, but is significantly below it. Thus its
position and rise speed do not correspond to the dash-dotted lines
for the front of the emerging tube shown in Figure 4.
In this sequence of images, a field line of a particular color corresponds to the
same field line carrying the same plasma.  A corresponding animated GIF movie showing the
evolution of the tracked field lines is also available from the electronic edition.

At the beginning of the flux emergence (see the $t=65$ image in Figure 8),
the field lines all show a normal configuration, i.e. they arch over the polarity
inversion line from the positive to the negative polarity region.
The $B_y$ components of these field lines at the midpoints are all positive.
Thus at this time, the O-point, which is the point of zero $B_y$ on the central
vertical line, is located below the black field line (the original tube axis).
With time the two polarity regions on the photosphere begin to show a significant
shearing motion along the neutral line \citep[e.g.][]{manchester2001,
manchestetal2004, fan2001, magara_longcope2003, magara2004, magara2006},
and the two polarity regions begin to
separate.  As a result the lower-lying black (original axial field line) and 
dark blue field lines are lengthened and begin to rotate (clockwise) toward an
inverse configuration, with their $B_y$ changing from positive to negative
at the midpoints (see Figure 8).  By time $t=90$ (see Figure 8), the black and dark blue
field lines have rotated to show a slight negative $B_y$ at their mid points,
while the upper yellow and red field lines are still having positive $B_y$ at
their midpoints.  This change of orientations of the lower (black and dark blue)
field lines from positive $B_y$ to negative $B_y$ causes an abrupt upward
shift of the zero point of $B_y$ along the center vertical line of $x=0$ and $y=0$,
i.e. an upward shift of the O-point
seen in Figure 4(a) (soon after $t=85$).  As a result the O-point
position shifts from below to above the original tube axis
(the black field line in Figures 7,8 and 9).  This apparent crossing of the O-point
position from below to above the original tube axis (as seen between $t=85$ and $t=90$
in Figure 4a) is not a true plasma rise motion, but is simply due to the change of
horizontal field line orientations in the atmosphere:  the original tube axis
(the black field line tracked in Figure 8) rotates from a forward into an inverse
configuration, with $B_y$ at its mid point changing from positive to negative,
and as a result the zero point of $B_y$ on the center vertical line of $x=0$ and $y=0$,
namely the O-point, propagates upward in position to a {\it different} field line above,
which is now aligned with the $x$ direction.

By time of about $t=90$, a significant counter-clockwise vortical or
rotational motion has set in within each of the two polarity flux
concentrations on the photosphere, reminiscent of sunspot rotations
\citep[e.g.][]{brownetal2003}.
The vortical motions can be seen from the footpoint motions of the field
lines in Figures 7,8, and 9 (and in the movie), and more clearly in Figure 10 which
shows snapshots of the $z$ component of the vorticity ${\omega}_z$ on the
photosphere overlaid with contours of $B_z$.
We find high intensity of positive ${\omega}_z$ (corresponding
to counter-clockwise vortical motion) centered on the peaks of the vertical
magnetic flux concentration of both polarities.
Figure 11 shows the evolution of the mean vertical vorticity $< \omega_z >$
averaged over the area of each polarity flux concentration where $B_z$ is above 
75\% of the peak $B_z$ value.
We find that the counter-clockwise vortical motion centered on each polarity
flux concentration, as represented by $< {\omega}_z > $, rises to its peak
rate at $t \approx 93$ and then persists at a relatively steady rate throughout
the subsequent evolution.
The footpoint vortical motions further twist up the emerged field lines, causing
continued clock-wise horizontal rotation of the blue and the higher yellow and
red field lines in the atmosphere as viewed from the top
(see the $t=100$, $t=110$, $t=118$, and $t=120$
panels in Figure 8 and the movie). With time, the yellow and then red field lines
are also progressively rotated into an inverse configuration, changing the sign
of $B_y$ from positive to negative at their mid points.
This causes the O-point position (the zero $B_y$ position on the center line of
$x=0$, $y=0$) to continue to propagate upward as seen in
Figure 4(a) after about $t=90$. Note again that this upward shift of the O-point
position from $t=86$ to about $t=102$ is not due to a true vertical rise of the plasma
as can be seen from the near zero $v_z$ during this period shown in Figure 4(b),
but is simply a propagation effect produced by the change
of horizontal orientations of the emerged field lines.

Starting from about $t=102$,
the plasma rise velocity $v_z$ measured at the O-point position also shows a
significant upward acceleration (see Figure 4(b)), which means that
the rapid ascent of the O-point after about $t=102$ (see Figure 4(a)) is a
combination of the upward propagation produced by the change of
the horizontal orientations of the field lines and an actual rise motion of the field lines.
At about $t=118$ (see Figure 8), the O-point position has reached $z=40.2$, which is
at this instant identified with the mid-point of the red field line. The red field line
has rotated to become aligned with the $x$-direction with zero $B_y$ at the middle and
is at this instant identified as the axial field line of the coronal flux rope.
The plasma rise speed $v_z$ measured at the O-point is about $2.6$ (or 16.9 km/s),
which continues to increase (see Figure 4(b)).  The field lines continue to
rotate horizontally. As a result the field line identified as the coronal axial
field line is continually changing.  By time $t=120$, the red field line has further rotated
to having a negative $B_y$ at its center, and the O-point has now shifted to a field
line (not shown in Figures 7, 8, and 9) above the red field line.

We also find that in conjunction with the horizontal rotation of the black and
blue field lines in the atmosphere due to shearing and twisting at their footpoints,
they develop central dips,
into which plasma drains, and the two side ``wings'' of each of these field lines
rise rapidly into the corona (see the $t=100$, $t=110$, $t=118$, $t=120$ images
in Figures 7 and 9).

The above examination of the tracked field lines indicate that the formation and
rise of the coronal flux rope is not a result of the bodily emergence of the
subsurface flux tube as a whole.
The bottom concave portions of the helical field lines in the
original emerging flux tube breaking through the surface remain largely trapped
at and below the photosphere layer due to the weight of the heavy plasma,
and the center of
the original tube axis ceases to rise after reaching about 2-3 pressure scale
heights above the photosphere.  
However, the shearing and vortical motions of the two polarity flux
concentrations at the photosphere continue to transport twist and magnetic energy
into the emerged field in the atmosphere
\citep[e.g][]{magara_longcope2003, manchestetal2004, magara2006}.
Our analysis in this paper shows that due to this vortical motion
at the footpoints, the emerged field lines near and just above the original axial 
field line (including the original axial field line itself) rotate horizontally in
the atmosphere, changing $B_y$ at their mid points from
positive to negative (i.e. from an original normal configuration to
an inverse configuration
with respect to the photosphere neutral line). This initiates an upward
{\it propagation} of the O-point in the emerged field in the central
$x=0$ cross-section simply because of the change of horizontal orientation
of the inner field lines, and not because of the rise of these field lines.
Later, the plasma rise speed measured at the O-point also
shows a sharp increase, contributing to the rapid ascent of the O-point into
coronal heights.  As a result, a coronal flux rope with an axis
identified with the field line instantaneously aligned with the $x$-direction
at its mid point (i.e. going through the O-point),
and with sigmoid shaped field lines threading under the axis with
an inverse configuration develops, and it is rising with increasing velocity as
the vortical motions at the footpoints of the flux rope continue to transport
magnetic energy and twist into the corona.
Note that the field line being identified as the axial field line of the
coronal flux rope is continually changing, due to the continued horizontal
rotation of the field lines in the atmosphere as their footpoints are being
twisted.

To quantify the transport of twist into the solar atmosphere due to both flux
emergence as well as the horizontal vortical motions at the photosphere, we have
computed the evolution of the relative magnetic helicity in the atmosphere above
the photosphere and the helicity flux through the photospheric boundary.
The relative magnetic helicity in the atmosphere above the photosphere $z=0$
is defined as \citep{berger_field1984, finn_antonsen1985}:
\begin{equation}
H_m = \int_{z > 0 } ({\bf A} + {\bf A}_p) \cdot
( {\bf B} - {\bf P} ) \; dV
\label{Hm}
\end{equation}
where ${\bf B}$ is the magnetic field in the unbounded half
space above $z=0$,
${\bf A}$ is the vector potential for ${\bf B}$,
${\bf P}$ is the reference potential field having the same
normal flux distribution as ${\bf B}$ on the $z=0$ boundary,
and ${\bf A}_p$ is the vector potential for ${\bf P}$.
The relative magnetic helicity $H_m$ is invariant with respect
to the gauges for ${\bf A}$ and ${\bf A}_p$.
Following \citet{devore2000}, we use the following specific
${\bf A}$ and ${\bf A}_p$ for computing $H_m$:
\begin{equation}
{\bf A}(x,y,z) = {\bf A}_p (x,y,0) - {\hat {\bf z}} \times
\int_0^{z} {\bf B} (x, y, z') \; dz' ,
\label{a}
\end{equation}
\begin{equation}
{\bf A}_p (x,y,z) = \nabla \times {\hat {\bf z}}
\int_z^{\infty} \phi (x,y,z') \; dz',
\label{ap}
\end{equation}
where,
\begin{equation}
\phi (x, y, z) = {1 \over 2 \pi} \int \int
{ B_z (x', y', 0) \over [ (x-x')^2 + (y-y')^2
+ z^2 ]^{1/2}} \; dx' dy'.
\label{phi}
\end{equation}
If we use the specific ${\bf A}$ and ${\bf A}_p$
given in equations (\ref{a}) through (\ref{phi}), it can
be shown that the expression
(\ref{Hm}) for the relative helicity $H_m$ reduces to simply
\begin{equation}
H_m = \int_{z > 0} {\bf A} \cdot {\bf B} \; dV.
\label{Hm_special}
\end{equation}
Note, the derivation of the relative magnetic helicity assumed
unbounded half space above the photosphere
\citep{berger_field1984,devore2000}.
We carry out the above integral within the simulation
domain above $z=0$ assuming that the field outside of the domain
is zero. Thus, our helicity integration only gives an accurate
result before the emerged field has expanded to reach the top
and side boundaries.
It has also been shown \citep{berger_field1984, devore2000}
that the rate of change of the relative helicity $H_m$ can
alternatively be computed by evaluating the helicity flux
through the $z=0$ boundary plane:
\begin{equation}
{d H_m \over dt} = -2 \int \int [({\bf A}_p \cdot {\bf v}) {\bf B}
- ({\bf A_p} \cdot {\bf B} ) {\bf v} ] \cdot {\hat {\bf z}} \; dx dy
\label{hmflux}
\end{equation}
where ${\bf A}_p$ is given by equations (\ref{ap})-(\ref{phi}).
The first term on the right-hand-side of equation (\ref{hmflux})
corresponds to the injection of helicity by horizontal motion
on the boundary, and the second term corresponds to injection
by direct emergence or submergence of twisted flux across
the boundary.
Note again, we carry out the above integration on the $z=0$ plane
within the domain, assuming zero flux through the $z=0$ plane
outside of the domain.
The resulting evolution of $d H_m/dt$ and $H_m$ for the emerged field
above $z=0$ is shown in Figure 12.  In the lower panel of Figure 12,
both $H_m$ computed from equation (\ref{Hm}) (solid line) and that
obtained by accumulation of helicity flux with time
$H_m = \int (dH_m /dt) \, dt$
(dash-dotted line) are shown.  The good agreement of the two
indicates good helicity conservation from our simulation.  The
deviation in the end is due to the fact that twisted magnetic flux
is exiting the top and side boundaries and is therefore not accounted
for in the integration of equation \ref{Hm}, while the time integration
of $dH_m / dt$ continues to give accurately the amount of helicity
that has been transported into the atmosphere through the boundary at
$z=0$.  The top panel of Figure 12 shows that the helicity flux
due to flux emergence is dominating for a brief period at the
beginning (from about $t=55$ to $t=80$), and then after about
$t=80$, horizontal shearing and rotational motions at the footpoints
of the emerged field become the main source of helicity flux.  
By time of about $t=90$, the rotational motions in the two polarity
flux concentrations have been established (see Figure 11) and they
provide a steady and dominant helicity flux into the atmosphere
in the subsequent evolution (top panel of Figure 12).
By time $t=120$, the helicity transported into the atmosphere
has reached about $- {\Phi}_{\rm tube}^2$, corresponding to one full
twist ($2 \pi$ rotation) of the flux ${\Phi}_{\rm tube}$ in the
original flux tube.

The rotational motions centered on the two polarity flux
concentrations are a manifestation of non-linear torsional
Alfv\'en waves propagating along the flux tube, transporting twist
from the tube's interior portion towards its expanded coronal portion.
Consider the quantity $\alpha \equiv
(\nabla \times {\bf B}) \cdot {\bf B} / B^2$, which gives a measure
of the local rate of twist of the magnetic field. For a force-free
magnetic field:
\begin{equation}
\nabla \times {\bf B} = \alpha {\bf B},
\label{fff}
\end{equation}
from which one can show that
\begin{equation}
{\bf B} \cdot \nabla \alpha = 0,
\label{eq_alpha}
\end{equation}
i.e. $\alpha$ is constant along each field line for
a field with zero Lorentz force.
If there is a gradient of $\alpha$ along a field line, then the
{\it thin} flux bundle (or tube) centered on the field line will
experience a net axial torque which drives torsional motion of
the flux tube \citep{longcope_klapper1997}.  The net axial
torque exerted on a thin flux tube segment $(l,l+dl)$,
where $l$ denotes the arc length along the thin tube, is given
by \citet[][eq. (20) in that paper]{longcope_welsch2000}:
\begin{equation}
\tau_l = {1 \over 16} a^4 B_l^2 {\partial \alpha \over \partial l}
dl
\label{tft_torque}
\end{equation}
where $\tau_l$ represents a torque along the axial direction $\hat{\bf l}$,
which will drive the tube segment to spin about its axis, $a$ denotes
the radius of the tube, $B_l$ denotes the axial field strength.
Equation (\ref{tft_torque}) states that the net axial torque $\tau_l$ for
a thin flux tube segment is directly proportional to the gradient of
$\alpha$ along the tube.
Note that in \citet{longcope_welsch2000}, the quantity $q$, defined as
the angle of field line rotation about the central axis per unit
length along the tube, is used in place of $\alpha$ to describe the twist.
For a thin flux tube, the relation between $\alpha$ and $q$ is $\alpha
= 2q$ \citep{longcope_klapper1997}.
Figure 13 shows the variation of $\alpha$ along three field lines: the
black field line shown in Figures 7, 8, and 9, corresponding to the original
tube axis, and its two neighboring blue field lines shown in
Figures 7, 8, and 9, at $t=118$.  We have plotted the $\alpha$ values along these
three field lines (using the same colors for the data points as those of
the corresponding field lines in Figures 7, 8, and 9) as a function of depth, from
their left ends to the left apices in the atmosphere, at $t=118$.
We can see that for all three neighboring field lines, the magnitude of
$\alpha$ decreases with height from the interior into the atmosphere.
The $\alpha$ is nearly constant along the atmosphere/corona
portions of the field lines, indicating
that they are near a force-free state, and the coronal value is
about a factor of 10 less than the peak magnitude of $\alpha$ in
the interior part of the flux tube.
The low $\alpha$ magnitude in the corona is due to the extreme
expansion and stretching of the magnetic field lines resulting from flux
emergence \citep{longcope_welsch2000}.
The lowered magnitude of $\alpha$ in the atmosphere establishes a
gradient of $\alpha$ along the flux tube from the interior to the
atmosphere, and produces a torque that drives the observed torsional
motion of the flux tube intersecting the photosphere.
The torsional motion will continue until the gradient of $\alpha$
along the tube is removed, and the coronal $\alpha$ equilibrates with
the interior $\alpha$ \citep{longcope_welsch2000}.

\subsection{Sigmoid fields and current sheet:}

The final 3D magnetic field in the atmosphere and corona obtained
from the simulation show a coronal flux rope structure whose
properties may qualitatively explain some basic observed
features of sigmoid active regions.
Similar to what was found in \citep{manchestetal2004, magara2004,
magara2006}, a current sheet has formed in the lower atmosphere.
The top panel of Figure 14 shows the current density $J$ in a
horizontal plane at height $z=10$ in the chromosphere. It
shows a horizontal cross-section of the current sheet.  The current
sheet has an over-all inverse-S shape (except the central zigzag).
Field lines traced through a few points along the current concentration
on $z=10$ are also plotted (for different perspectives) in the
two lower panels of Figure 14.
We find that all the field lines going through the current sheet show
an inverse-S shape when viewed from the top (as illustrated in the lower
right panel of Figure 14), similar to what has been found in the simulations of
\citet{magara2004, magara2006} and consistent with the theoretical
prediction of \citet{titov_demoulin1999} and \citet{low_berger2003}.
One can argue that the heating
associated with the current sheet may cause these inverse-S shaped field
lines to preferentially brighten up in soft X-ray, giving rise to the
observed X-ray sigmoid loops in an active region.
It can also be seen that some of the field lines undergo sharp dips at the
center portion of the current sheet (see the red field lines in the
lower left panel of Figure 14).  Reconnections in the current sheet 
are disengaging the heavy mass trapped in the dips, allowing for the
flux rope to rise further into the corona with increasing speed.

We find that the inverse-S shaped current sheet begins to
develop at the time of about $t=102$, being initially confined to
a few pressure scale-heights above the photosphere.  With time the current
sheet extends upward, and by time $t=120$, it has extended to the base
of the corona. Growing reconnections in the current sheet due to the
numerical diffusion inherent in the code are expected
to aid the upward acceleration of the center of the
flux rope seen after about $t=102$ (Figure 4b).  However, the ultimate
driver of the rise of the flux rope and the development of
the current sheet is the continued vortical motions at the footpoints
of the emerged field lines, which transport twist and magnetic energy
into the atmosphere, causing an excess of the upward Lorentz force that
drives the acceleration.

\section{Discussions}
We have presented a 3D MHD simulation of the emergence of a twisted
$\Omega$-shaped flux tube from the top layer of the
solar interior into the solar atmosphere and the corona. Compared to
some of the previous simulations \citep[e.g.][]{fan2001, archontisetal2004,
manchestetal2004}, the subsurface rising flux tube is more sharply
arched such that the emerging segment first breaking through the photosphere
contains less than one full wind of twist.  In this simulation, we find
that after an initial stage of flux emergence,
during which the two polarity flux concentrations on the photosphere
undergo a shearing motion and become
separated, a prominent rotational/vortical motion develops within
each polarity, reminiscent of sunspot rotations.
This rotational motion persists throughout the subsequent evolution and
steadily transports twist or magnetic helicity into the atmosphere.

Similar to what has been found in many previous investigations
\citep[e.g.][]{fan2001,archontisetal2004,manchestetal2004,magara2006},
the twisted flux tube is unable to rise bodily into the corona as a whole.
While the upper parts of the winding field lines of the twisted flux tube
expand into the atmosphere and the corona, the center of the original axial
field line of the tube ceases to rise at about 2-3 pressure scale heights
above the photosphere and the U-shaped parts of the helical field lines
largely remain anchored at and below the photospheric layer.
However, we find that the initial shearing
and the subsequent rotational motions of the two polarity flux concentrations
twist up the inner field lines (near and include the original tube axis)
of the emerged field, causing them to rotate in the atmosphere from
their initial normal configuration (having a positive $B_y$ at the mid point,
or arching over the polarity inversion line from the positive to the negative
polarity),
into an inverse configuration (having a negative $B_y$ at the mid point,
directed from the negative polarity to the positive polarity over the neutral
line).  This change of the horizontal orientations of the emerged field lines
causes an upward propagation of the O-point (the zero-crossing point of
$B_y$ along the center vertical line of $x=0$ and $y=0$) in the emerged field
in the central vertical cross-section ($x=0$).
With the continued twisting of the emerged field lines
by the vortical motions at the footpoints, which transports twist
and magnetic energy into the corona, the actual plasma motion
measured at the O-point also starts to accelerate upwards. Thus,
the rapid ascent of the O-point is a combination of the upward propagation
produced by the change of the horizontal orientations of the field lines
and the actual rise of the field lines.
As a result, a coronal flux rope with
an axial field line identified with the field line instantaneously aligned
with the $x$-direction at its mid point
(i.e. going through the O-point), and
with sigmoid shaped field lines threading under the axis with an inverse
configuration, is formed and rises with increasing speed as the vortical
motions at the footpoints continue.  Some of the sigmoid shaped field lines
threading under the axis have also develop sharp dips as they are
being twisted at the footpoints, with plasma draining into the dips
and with the two side lobes rising rapidly into the
corona.  A current sheet forms in the lower atmosphere with field lines
going through it all showing sigmoid
shapes.  Heating in the current sheet may therefore cause these field
lines to brighten up in X-ray as the observed sigmoid loops in active
regions.  Field lines going through the center portion of the current sheet
undergo sharp dips. Reconnections in the current sheet allow for removal
of the heavy plasma trapped in the dips and hence enhance the upward
acceleration of the coronal flux rope.

The formation of a coronal flux rope with sigmoid shaped core fields
and with a current sheet in the lower atmosphere as a result of the
emergence of a twisted flux tube from the interior into the solar atmosphere
has been found to be a very general result
\citep[e.g.][]{manchestetal2004, magara2004, magara2006,
archontis_toeroek2008}.  The onset of a rotational/vortical
motion centered on the two polarity flux concentrations soon after they
become separated has also been found in \citet{magara2006}.
Our work explicitly demonstrates how the vortical motions of the
footpoints twist up the inner emerged field lines and rotate them
into an inverse configuration (Figures 7, 8, and 9), leading to the
development and upward ascent of the coronal flux rope structure.
We also show that the rotational motions of the two polarity flux
concentrations on the photosphere are a manifestation of nonlinear
torsional Alfv\'en waves propagating along the flux tube, consistent with
what has been predicted by an idealized analytical model
of \citet{longcope_welsch2000}.
Due to the rapid stretching of the emerged magnetic field in the atmosphere
and corona, the magnitude of $\alpha = {\bf J} \cdot {\bf B} / B^2$ along
the coronal field lines, which is a measure of the rate of twist per unit
length, drastically decreases.
As a result, a gradient of the rate of
twist is established along the flux tube from the interior into the atmosphere
with the interior portion of the flux tube having a much higher
rate of twist.
This gradient in the rate of twist drives torsional waves along the flux tube,
transporting twist from the interior highly twisted portion into the expanded
coronal portion \citep{longcope_welsch2000}.
Thus the rotational motion will continue until the coronal $\alpha$
equilibrates with the interior $\alpha$ along the field lines. 
The time scale for establishing the equilibrium is on the order of
the Alfv\'en transient time along the interior flux tube, which means
that the rotational motion can persists for a few days after the initial
emergence.  Depending on how great the stretching of the emerged field lines
is compared to the total length of the interior flux tube, 
it is conceivable that the final equilibrated $\alpha$ value is
of a significantly smaller
magnitude compared to the initial $\alpha$ of the subsurface flux tube
rising to the photosphere.
This may provide an explanation for the significantly lower $\alpha$
values measured for the majority of the solar active regions compared to
the twist rate needed for a buoyant flux tube to rise cohesively in the
solar interior.

Sunspot rotations have been observed in many events preceding X-ray sigmoid
brightening and the onset of eruptive flares \citep[e.g][]{brownetal2003,
zhangetal2008}.  Our results suggest that these rotational motions
are due to non-linear torsional Alfv\'en waves naturally occurring
during the emergence of a twisted flux tube, and is
an important means whereby twist is transported from the interior into
the corona, driving the development of a coronal flux rope as a precursor
structure for solar eruptions.
The horizontal vortical motion and its subsurface extension corresponding
to the torsional Alfv\'en waves may be detectable by local helioseismology
techniques.

\acknowledgements
I thank B.~C. Low and an anonymous referee for helpful comments on
the manuscript.  NCAR is sponsored by the National Science Foundation.
This work is also supported in part by the NASA Heliophysics Guest
Investigator Grant NNG07EK66I.  The numerical simulations have
been carried out on the SGI ALTIX system and IBM POWER5+ system
at NASA Advanced Supercomputing Division under project GID s0869,
and also on NCAR's IBM POWER6 system.

\clearpage
\begin{figure}
\epsscale{0.5}
\plotone{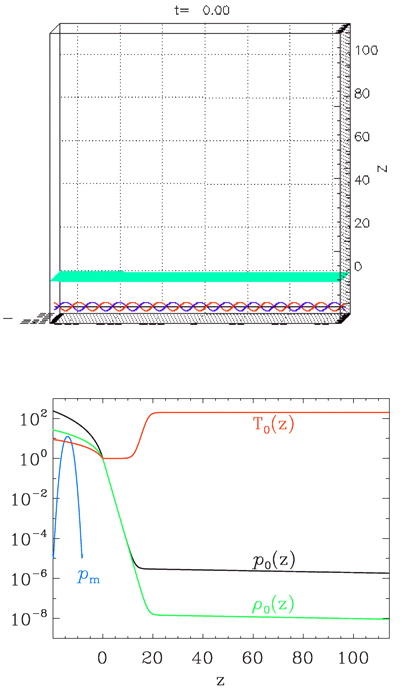}
\figcaption[f1.png]{The initial setup of the simulation. The top
panel shows the simulation domain, where the green plane denotes the
photosphere and the configuration of the initial flux tube is illustrated
with 3 traced field lines in the flux tube.  The bottom panel shows the
vertical profiles of plasma pressure, density and temperature of the
field-free static atmosphere, and also the profile of the magnetic
pressure through the embedded flux tube along the center line at
$(x,y) = (0,0)$.}
\end{figure}

\begin{figure}
\epsscale{1.}
\plotone{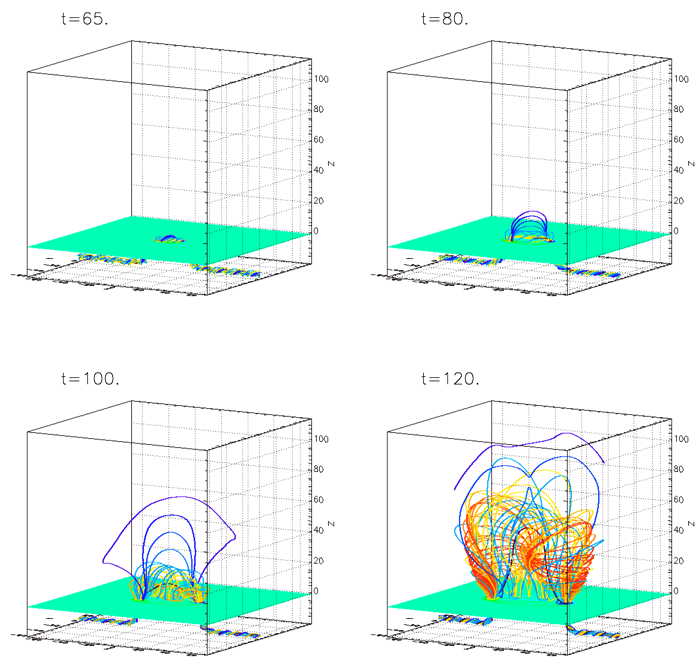}
\figcaption[f2.png]{Snapshots of the 3D magnetic 
field during the flux emergence into the atmosphere. Note that field lines
in the different snapshots are not the same field lines carrying the same
plasma - they are simply field lines selectively drawn in each instance
to reflect the topological structure of the 3D field.
The black field line shown in each of the panels
is the field line that is drawn from the O-point of the transverse field
in the central cross-section (at $x=0$), and it corresponds to a different
field line in each time.}
\end{figure}

\begin{figure}
\epsscale{1.}
\plotone{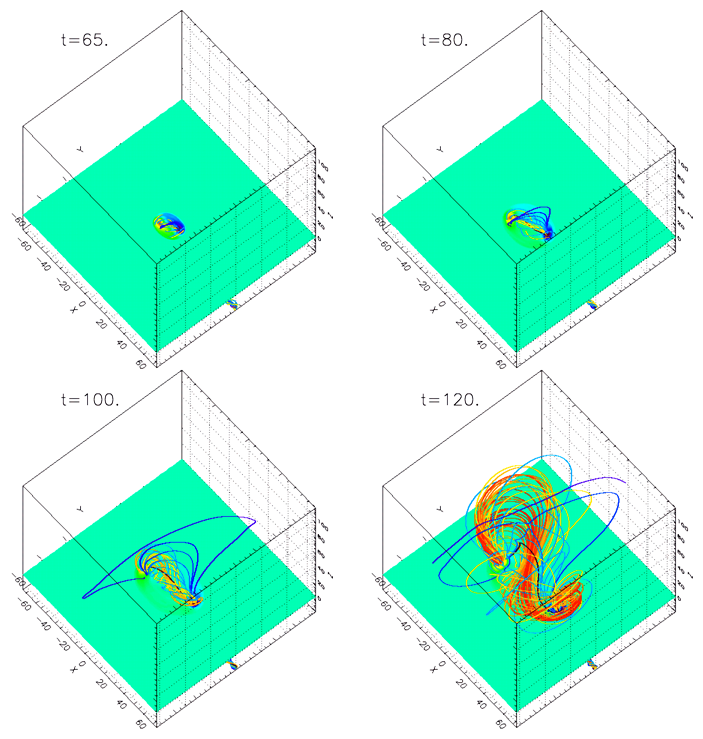}
\figcaption[f3.png]{Same as Figure 2 but viewed from
a different perspective.}
\end{figure}

\begin{figure}
\epsscale{0.6}
\plotone{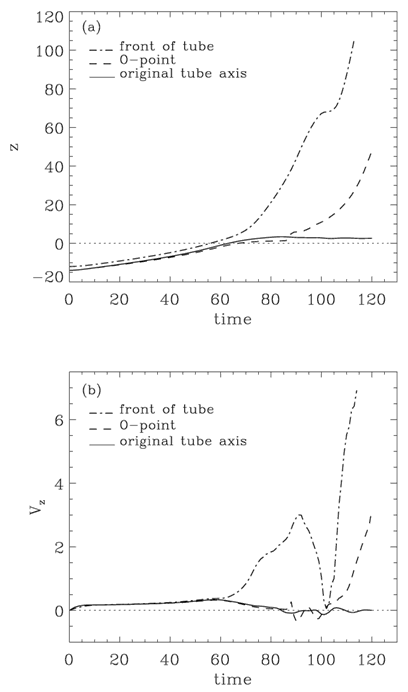}
\figcaption[f4.png]{The evolution of the height (a) and vertical
velocity (b) for 3 different points within the central vertical plane
of $x=0$: (1) the front of the emerging flux tube (dash-dotted line),
(2) the Lagrangian position of the original tube axis (solid line), and
(3) the O-point of the transverse magnetic field in the central
cross-section (at $x=0$), which is the point where $B_y=0$
on the center vertical line of $x=0$ and $y=0$.
The mid-points of the black field lines shown in Figure 2 and 3
correspond to the O-point positions at those instances.
}
\end{figure}

\begin{figure}
\epsscale{0.6}
\plotone{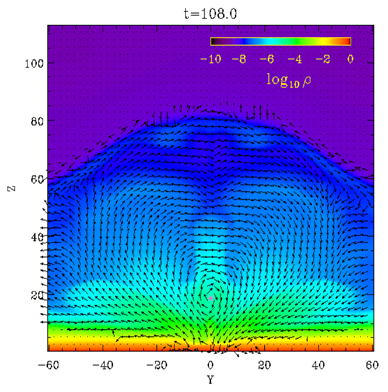}
\figcaption[f5.png]{A snapshot of the central vertical plane of $x=0$
at $t=108$, with arrows showing the direction of the transverse magnetic field
(lengths of the arrows are all the same, normalized by the transverse magnetic
field strength for all points where the transverse field strength is non-zero),
overlaid with a color image showing the density in log scale. Only the part above
the photosphere is shown. The O-point of
the transverse magnetic field is marked by the pink `*' point}
\end{figure}

\begin{figure}
\epsscale{0.7}
\plotone{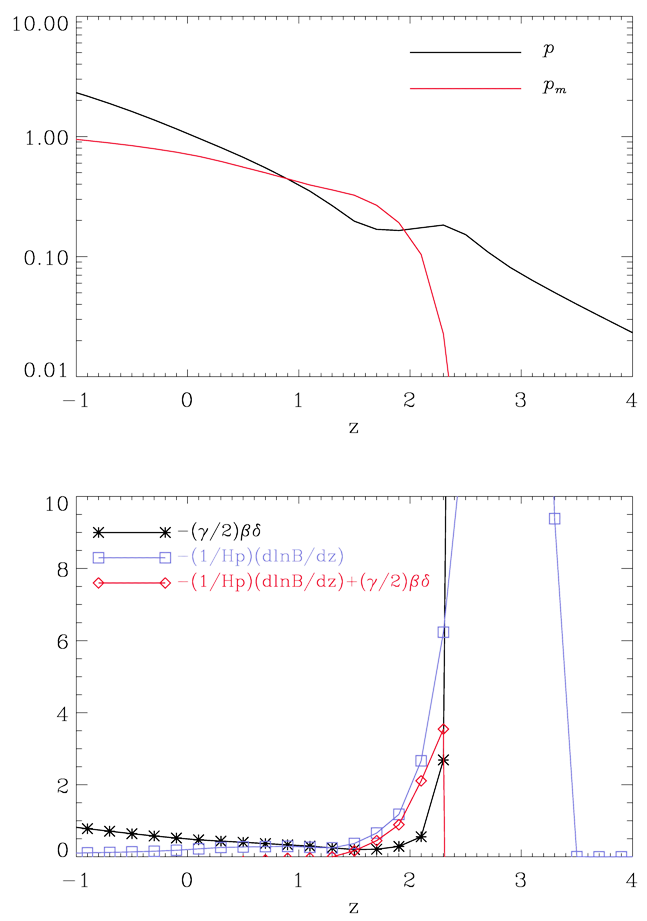}
\figcaption[f6.png]{Vertical profiles of several quantities
near the photosphere along the center line of $(x,y)=(0,0)$, at $t=60$.
It shows the conditions of the emerging field entering the photosphere 
when it becomes unstable to the onset of the magnetic buoyancy instability
(see discussion in the text)}
\end{figure}

\begin{figure}
\epsscale{0.8}
\plotone{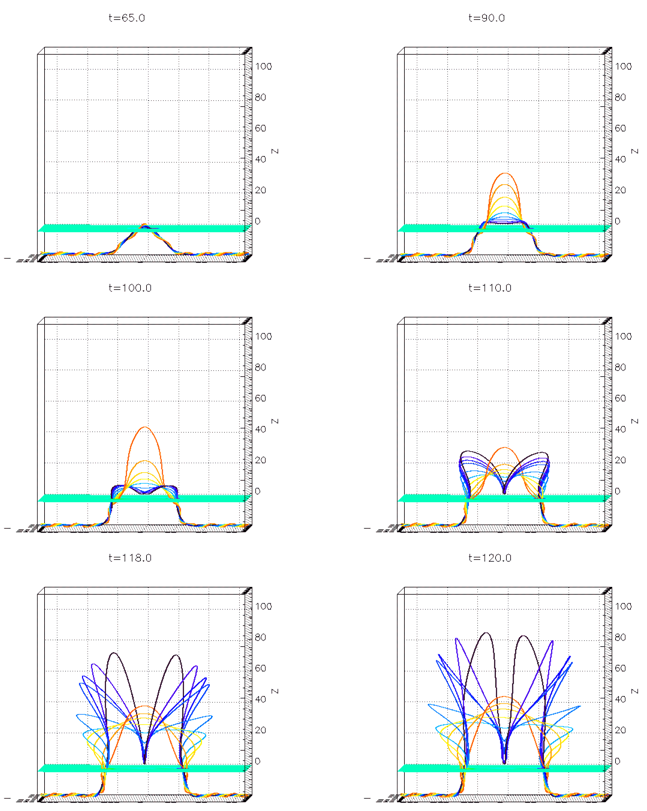}
\figcaption[f7.png]{3D evolution of a set of tracked field lines
as they are being twisted up by the shear and rotation motions
at their footpoints on the photosphere.  In this sequence of images,
a field line of a particular color corresponds to the same field line
carrying the same plasma.  The black field line corresponds to the original
tube axis, and all the other field lines have their mid points (at $x=0$, $y=0$)
above the mid point of the black field line (reddish field lines have mid
points above bluish field lines).  An animated GIF movie is also available
in the electronic version.}
\end{figure}

\begin{figure}
\epsscale{0.8}
\plotone{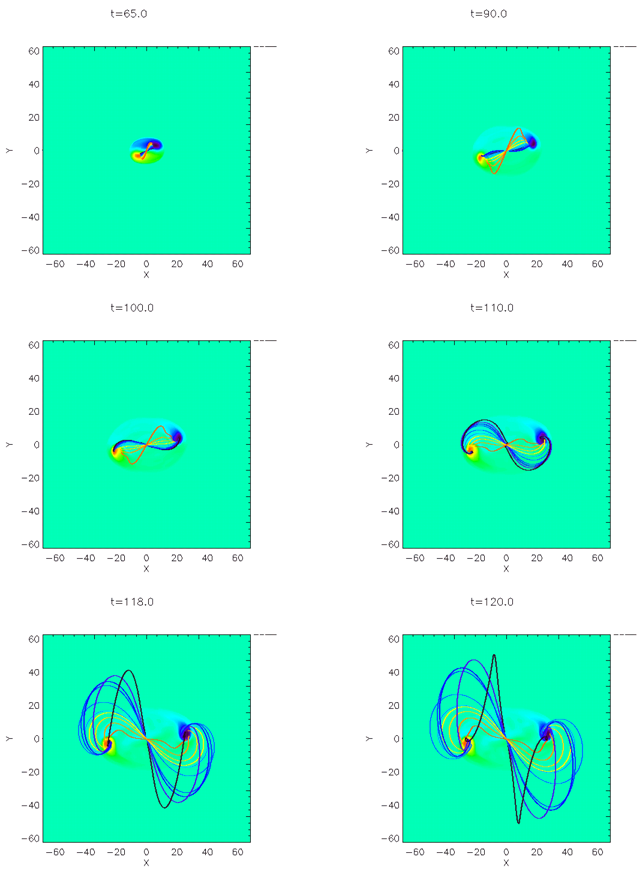}
\figcaption[f8.png]{Same as Figure 7, but is viewed from
the top}
\end{figure}

\begin{figure}
\epsscale{0.8}
\plotone{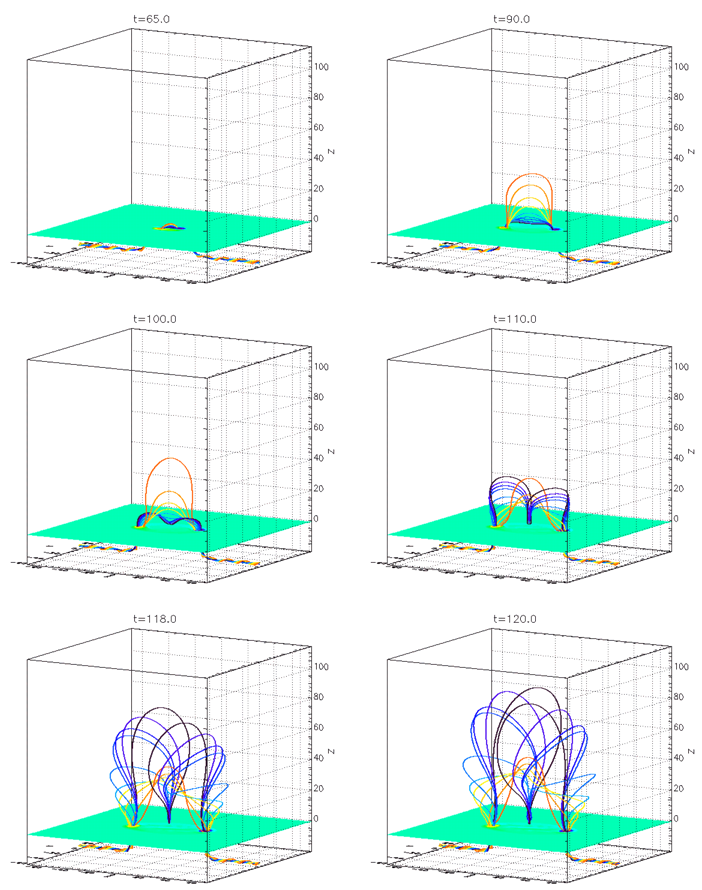}
\figcaption[f9.png]{Same as Figure 7, but
a different side view perspective.}
\end{figure}

\begin{figure}
\epsscale{0.5}
\plotone{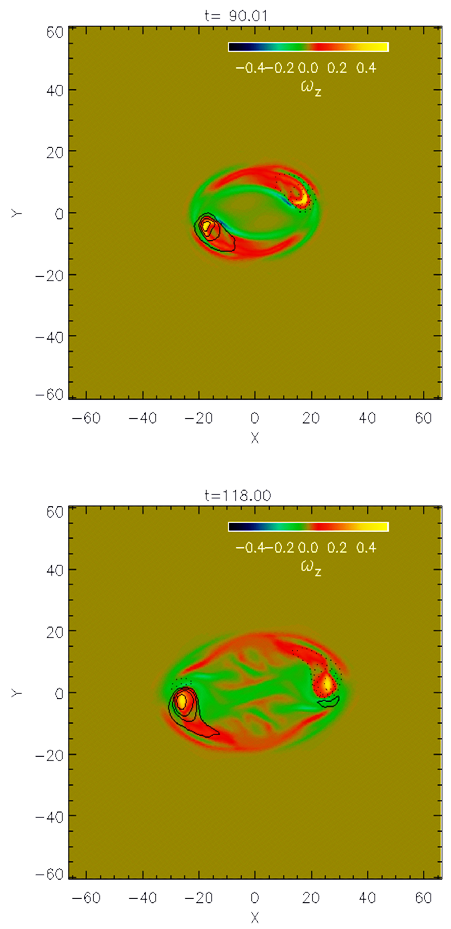}
\figcaption[f10.png]{Snapshots of the $z$ component of the vorticity
${\omega}_z$ on the photosphere overlaid with contours of $B_z$
with solid (dashed) contours representing positive (negative) $B_z$.
The images show counter-clockwise vortical motion centered on the peaks
of the vertical flux concentration of both polarities.}
\end{figure}

\begin{figure}
\epsscale{0.8}
\plotone{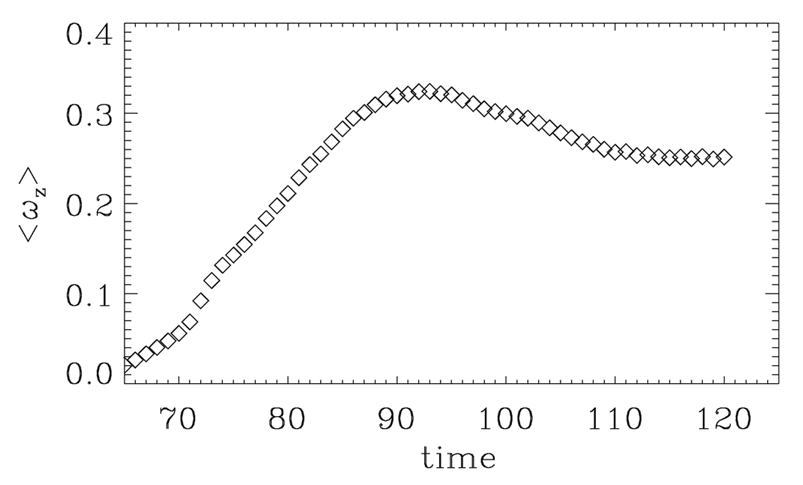}
\figcaption[f11.png] {The evolution of the mean vertical vorticity
$< \omega_z >$ averaged over the area of each polarity flux concentration
where $B_z$ is above 75\% of the peak $B_z$ value.}
\end{figure}

\begin{figure}
\epsscale{0.7}
\plotone{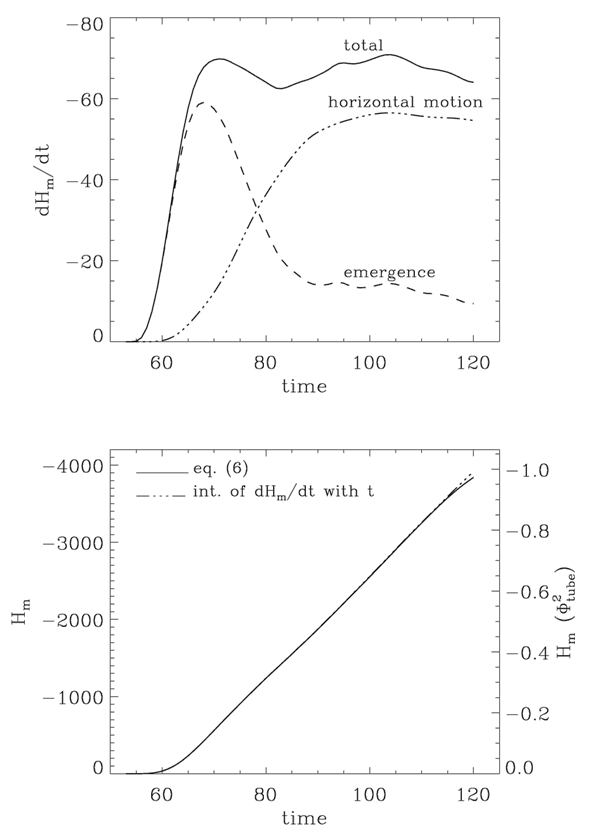}
\figcaption[f12.png]{Top panel shows the helicity injection into the
atmosphere $d H_m / dt$ due to horizontal motions on the photosphere
(dash-dotted line), vertical emergence of flux (dashed line), and the sum
of the two (solid line).  The bottom panel shows the evolution of the
relative magnetic helicity $H_m$ of the emerged field computed from
equation (6) (solid line) and through time integration of the
total $d H_m / dt$ (dash-dotted line).}
\end{figure}

\begin{figure}
\epsscale{0.7}
\plotone{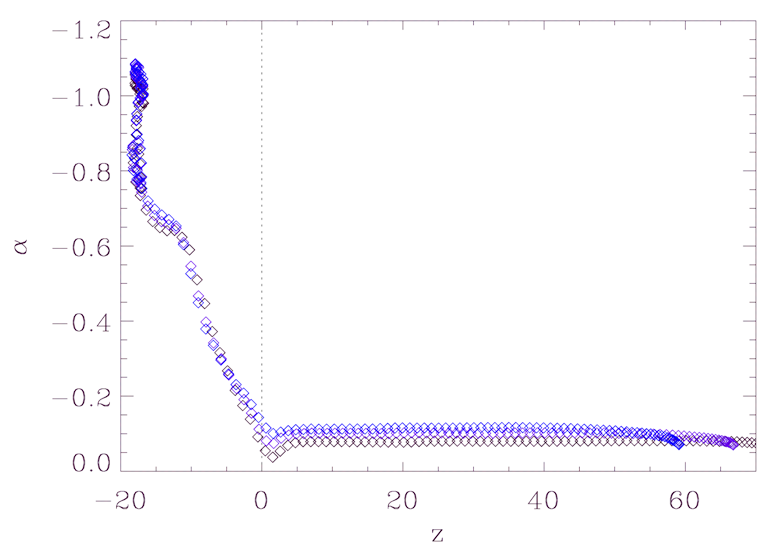}
\figcaption[f13.png]{The variation of $\alpha \equiv
(\nabla \times {\bf B}) \cdot {\bf B} / B^2$ as a function of $z$ along
three field lines: the black field line shown in Figures 7, 8, and 9, which is
the original tube axis, and its two neighboring blue field lines shown
in Figures 7, 8, and 9, at time $t=118$. The $\alpha$ values
are plotted along these three field lines (using the same colors for the
data points as those of the corresponding field lines in Figures 7, 8, and 9) as a
function of depth, from their left ends to the left apices in the atmosphere.}
\end{figure}

\begin{figure}
\epsscale{1.}
\plotone{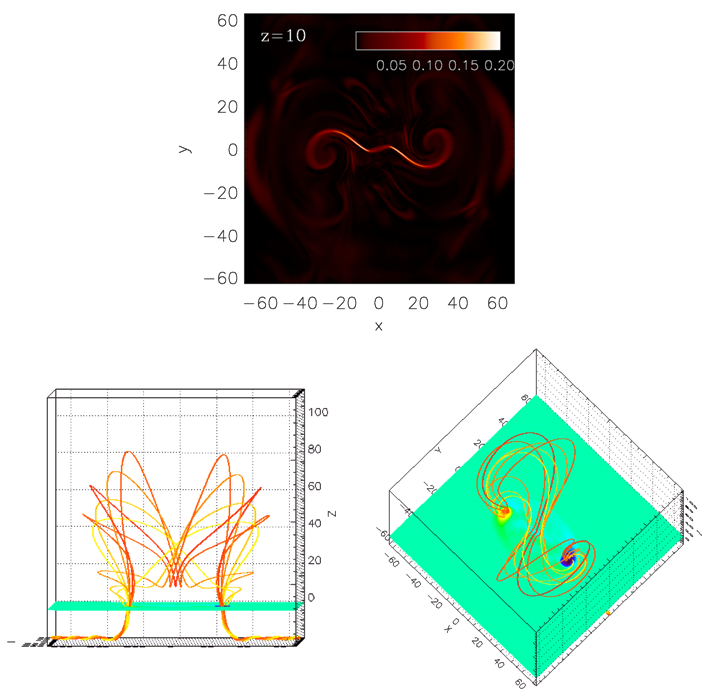}
\figcaption[f14.png]{The top panel shows the current density
$J = \nabla \times {\bf B}$ in a horizontal plane at height $z=10$ in
the chromosphere. The two lower panels show two perspective
views of a set of 3D field lines traced from a few points along the
current concentration shown in the top panel.}
\end{figure}

\end{document}